\documentclass[mask, anonymous]{egpubl}
\usepackage{eurovis2021}
\usepackage{multicol}
\usepackage{multirow}
\usepackage{amsmath}
\DeclareMathOperator*{\argmax}{arg\,max}

\EuroVisShort  

\usepackage[T1]{fontenc}
\usepackage{dfadobe}  

\usepackage{cite}  

\usepackage{xcolor}
\usepackage{enumitem}

\newif\ifnotes
\notestrue

\BibtexOrBiblatex
\electronicVersion
\PrintedOrElectronic
\ifpdf \usepackage[pdftex]{graphicx} \pdfcompresslevel=9
\else \usepackage[dvips]{graphicx} \fi

\usepackage{egweblnk}

\title[EG \LaTeX\ Author Guidelines]%
      {Guided Hyperparameter Tuning \\ Through Visualization and Inference}

\author[Hyekang Joo \& Calvin Bao \& Ishan Sen \& Furong Huang \& Leilani Battle]
{\parbox{\textwidth}{\centering 
    Hyekang Joo\thanks{The two authors contributed equally to this research}\orcid{0000-0002-6387-5686},
    Calvin Bao\footnotemark[1]\orcid{0000-0002-7063-221X},
    Ishan Sen\orcid{0000-0002-4888-7785},
    Furong Huang,
    and Leilani Battle
    }
        \\
{\parbox{\textwidth}{\centering University of Maryland, College Park, MD, USA}
}
}

\begin{document}

\maketitle
\begin{abstract}
For deep learning practitioners, hyperparameter tuning for optimizing model performance can be a computationally expensive task. Though visualization can help practitioners relate hyperparameter settings to overall model performance, significant manual inspection is still required to guide the hyperparameter settings in the next batch of experiments. In response, we present a streamlined visualization system enabling deep learning practitioners to more efficiently explore, tune, and optimize hyperparameters in a batch of experiments. A key idea is to directly suggest more optimal hyperparameter values using a predictive mechanism. We then integrate this mechanism with current visualization practices for deep learning. Moreover, an analysis on the variance in a selected performance metric in the context of the model hyperparameters shows the impact that certain hyperparameters have on the performance metric. 
We evaluate the tool with a user study on deep learning model builders, finding that our participants have little issue adopting the tool and working with it as part of their workflow.


\end{abstract}  
\section{Introduction}

Hyperparameters play a crucial role in machine learning because they directly control the behaviors of learning algorithms and have a significant effect on the performance of the associated models \cite{Goodfellow-et-al-2016}.
We refer to the process of finding hyperparameter configurations that results in higher performance as \emph{hyperparameter tuning}.
Specifically, given a hyperparameter configuration $h \in \mathcal{H}$, performance metric $p \in \mathcal{P}$ and a learning algorithm $f: \mathcal{H} \rightarrow \mathcal{P}$, we aim to estimate $\argmax{_{h \in \mathcal{H}}} f(h) $ that results in a better performance.
This process commonly involves iteratively selecting a hyperparameter configuration, training a model with the configuration, and evaluating the performance with established metrics \cite{tuningsurvey2020}. 

However, as deep learning algorithms and models become increasingly more complex, hyperparameter selection becomes more unwieldy a task for users.
As this process is computationally expensive \cite{Goodfellow-et-al-2016}, it becomes crucial to narrow down the search space; as most performance variance can be attributed to just a few hyperparameters \cite{pmlr-v32-hutter14}, identifying the hyperparameters that have the greatest influence on the performance metric, and tuning those hyperparameters, is of utmost importance.
Though manual exploration allows users to gain a deeper understanding for the influence of different hyperparameters, enumerating all possible hyperparameter configurations and checking their performance impact is not tractable, requiring a more efficient and automated approach \cite{Goodfellow-et-al-2016, snoek2012practical, park2020hypertendril}. 

\begin{figure*} [hbt!]
    \centering
    \includegraphics[width=\textwidth,height=6cm]{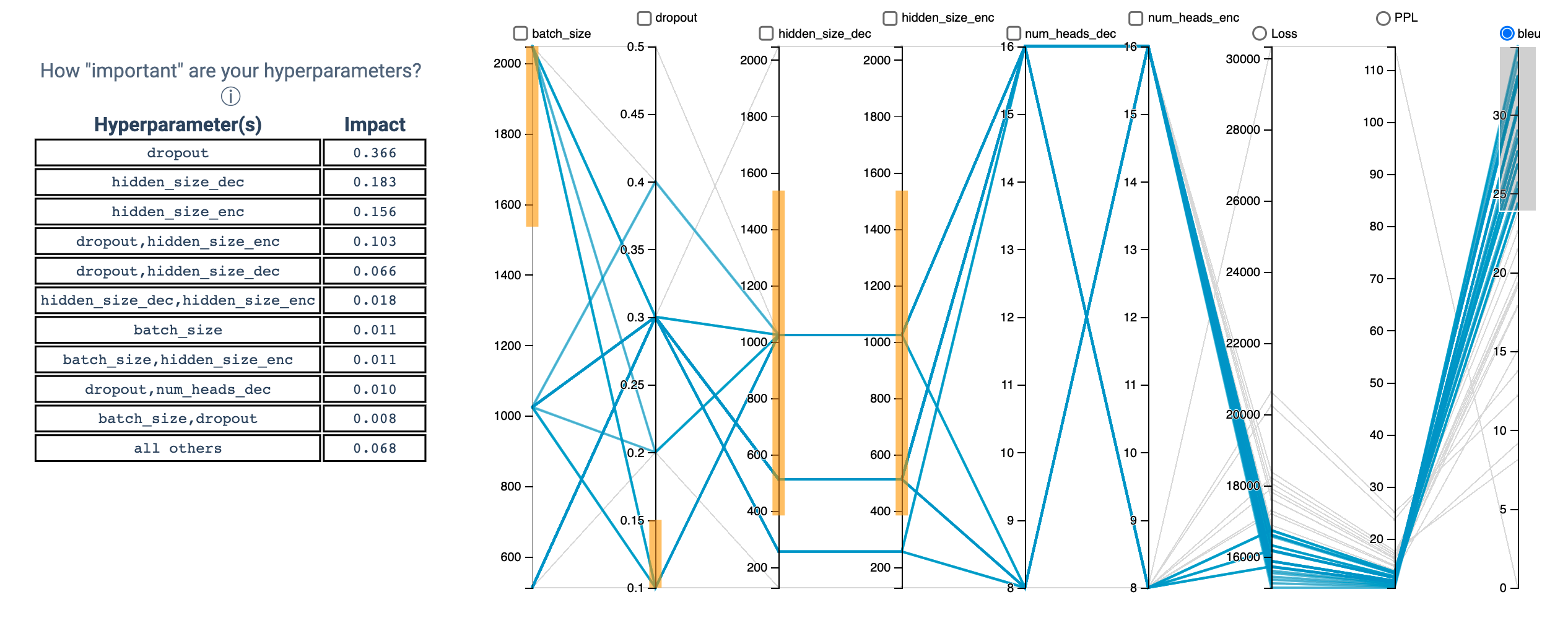}
    \caption{
    Parallel Coordinate Plot and supporting descriptive visualizations: Importance score table is shown on the left; Specific importance score is shown by selecting a set of hyperparameters (six square boxes present atop the first six axes) and one performance metric (three radio buttons atop the three rightmost axes); Each model is represented as a single blue polyline in the parallel coordinates visualization on the right; The axes on the left side (under six square boxes) of the parallel coordinates plot represent hyperparameters (\textit{e.g.}, \textit{batch\_size}, \textit{dropout}, \textit{etc.}), and those on the right side (under three radio buttons) are performance metrics (\textit{e.g.}, \textit{PPL}, \textit{bleu}, \textit{etc.}); The orange represents a range of the predicted optimal values for the hyperparameter. 
    }
    \label{fig:parallelplot}
\end{figure*}

Many algorithms have been developed to identify models that optimize a particular evaluation metric~\cite{NIPS2011_4443, snoek2012practical, BayesianVictoria2020,RandomSearch2012,Jaderberg2017PopulationBT, PopBasedLiSpyra2019}, but there is no guarantee that heuristics-based optimization will actually provide optimal hyperparameter values \cite{Goodfellow-et-al-2016}. In addition, these algorithms do not inform the user of which hyperparameters hold more weight when it comes to improving overall model performance. Instead of recommending specific hyperparameter values, we encourage a guided, user-driven approach to hyperparameter optimization by including suggested ranges over hyperparameter values and a ranking of important hyperparameters. We integrate these into a streamlined visual analytics tool (see \autoref{fig:parallelplot}).

In this paper, we make the following contributions:
\begin{itemize}[nosep]
  \item a model for predicting high-performing hyperparameter ranges, to guide the user's decisions in building future models
  \item a model for ranking hyperparameters by their impact on a selected performance metric (\textit{e.g.}, \cite{park2020hypertendril})
  \item a streamlined interface integrating our models directly with current best practices in visualizing deep learning models (\textit{e.g.}, \cite{park2020hypertendril,tensorflow2015-whitepaper})
  \item a preliminary study with four users; our results show that users find our automated features easy to use, and more effective than existing hyperparameter tuning techniques
\end{itemize}


\section{Related Work}
\subsection{Visual Analytics Systems for Machine Learning}
Recently, there has been a surge in visual analytics systems that enable analysts to better understand ML models. A category of systems interactively tweak feature values to see how a prediction responds \cite{Krause2016, amershi2015modeltracker}, which aid in understanding model sensitivity toward features. 

Another category includes systems such as \textit{ModelTracker} \cite{amershi2015modeltracker}, \textit{CometML} \cite{cometml}, and \textit{TensorFlow} \cite{tensorflow2015-whitepaper}, which encourage an "informed approach" to model building. They allow users to monitor model performance by displaying the relationship between the hyperparameters used during training and the performance of the model. In doing so, users are able to easily identify issues within the space and see the impact on model performance. Our approach falls under the "informed approach" category, and we build upon existing features to facilitate a guided and interpretable analysis of models. 

\subsection{Hyperparameter-driven Visual Analytics Systems}
Existing visual analytics systems that fall under the category of hyperparameter-driven comparisons include \textit{CometML} \cite{cometml}, \textit{TensorFlow} \cite{tensorflow2015-whitepaper}, \textit{HyperTuner} \cite{LiConvertino2018}, and \textit{HyperTendril} \cite{park2020hypertendril}. \textit{CometML}, \textit{TensorFlow}, and \textit{HyperTuner} primarily distinguish models based on unique hyperparameter configurations, and construct visualizations that focus on model performance in order to ease model comparison and selection for the user. 
Meanwhile, \textit{HyperTendril} takes this task a step further by leveraging a quantifiable metric known as \textbf{hyperparameter importance}, or a hyperparameter's degree of impact on a selected performance metric, using an efficient framework for functional ANOVA (fANOVA) analysis \cite{pmlr-v32-hutter14}. 

While recent work such as \textit{HyperTendril} has provided an individual quantifiable importance score to the user by embedding it in their dashboard, we further develop the idea of a guided approach to hyperparameter optimization by proposing a predictive approach, explained in \autoref{sec:model:random-forest}, in tandem with the aforementioned hyperparameter importance.
We provide a visual representation of our predictive model to the user by highlighting bounds for each hyperparameter in a parallel coordinates plot that jointly contribute to maximal performance metrics. This approach enables us to refine the hyperparameter value space for the next batch of experiments.

\begin{figure}[htbp]
    \centering
    \includegraphics[width=\linewidth]{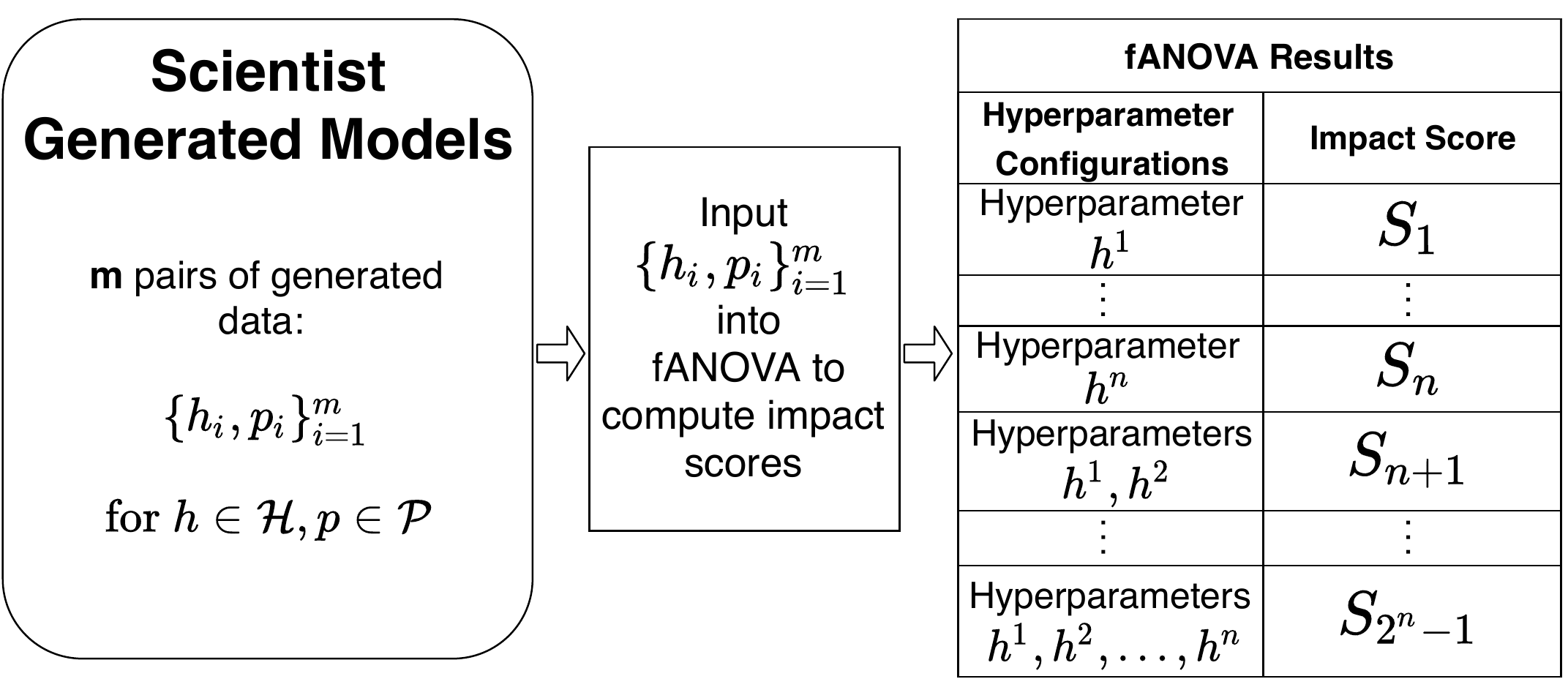}
    \caption{Functional Analysis of Variance Pipeline, where each $h$ is a configuration of hyperparameters $h^1 ... h^n$, $p$ is a specific performance metric, and S is an impact score ($\sum_i{S_i} = 1$).}
    \label{fig:fanova_pipeline}
\end{figure}

\section{Guided Hyperparameter Tuning}\label{sec:predictive}

This section describes the implementation of two features that are leveraged for visual guidance in hyperparameter tuning. It also details the initialization of two dashboards, purposed for different deep learning tasks, specifically, image classification and machine translation. These dashboards were evaluated through a user study to gauge the effectiveness of these features in the context of a visual analytics system.


\subsection{Hyperparameter Impact Analysis}
\label{sec:model:fanova}
\begin{figure*}[htbp!]
  \centering
  \includegraphics[width=0.9\linewidth,height=4.5cm]{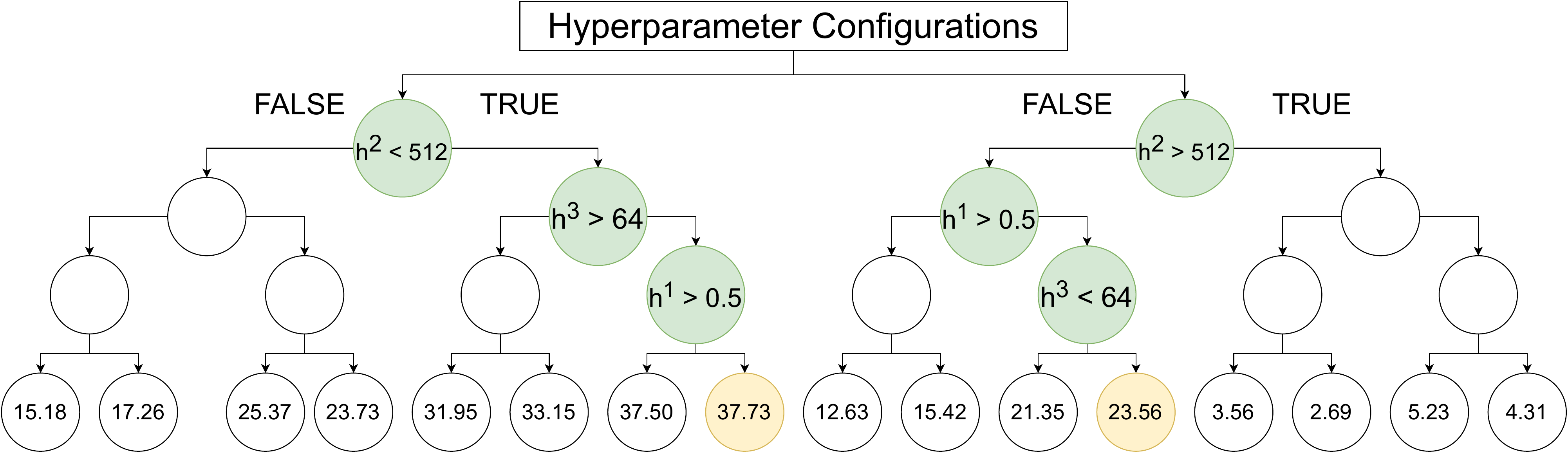}
  \caption{Visual representation of how predictive bounds are determined for the dashboard, after fitting $\mathcal{H}\rightarrow\mathcal{P}$. Each tree is a simplified version of true trees, and the green indication on each tree represents the path to the optimal leaf of that tree (colored yellow). Tight bounds are determined by observing decisions made in each tree. An example of tight bounds for this forest of 2 trees are $h^1>0.5$, $256 < h^2 < 512$, and $h^3>64$.}
  \label{fig:rf}
\end{figure*}
To measure how hyperparameter configurations impact a performance metric, functional analysis of variance (fANOVA) \cite{pmlr-v32-hutter14} is used. An implementation of fANOVA can be found as part of the AutoML toolkit\footnote{\url{https://github.com/automl}}, and it allows us to analyze what proportion of the performance variance can be explained by a single hyperparameter, or combinations thereof, in linear time. We denote the proportion of variance explained by the hyperparameter as an impact score, where a higher score indicates more variance being explained, and the sum of these scores adds up to 1. This is a simple yet effective method of understanding the relationship between the hyperparameter space and performance of the model. The integration of fANOVA in our dashboard can be seen in \autoref{fig:fanova_pipeline}. 

\subsection{Learning Procedure for $\mathcal{H}\rightarrow\mathcal{P}$ } 
\label{sec:model:random-forest}

We formulate a regression problem fitting the model's quantitative performance metric to the hyperparameter configuration. We sought to leverage efficient and interpretable models. The benefit of interpretability is that we are able to deconstruct the model into parts, which we can visually display to the user and potentially use as part of the visual guiding process. 

One such modeling strategy is multiple linear regression, with which we could extract the weights of each feature (the hyperparameter) for showing to the end-user; however, this resulted in a poor-performing model. Another strategy is random forest regression \cite{BreimanRandomForest} which could account for the non-linear relationship between hyperparameter and performance metric. This method enables us to deconstruct the model into its components and presents us with a guide for optimal hyperparameter ranges accordingly.

More specifically, this enables us to extract sets of logical rules from nodes found in the decision trees from the random forest: the highest yielding path in a tree is selected, backtracked on to identify logical decisions made at each node ($h^i<0.45,h^{i+1}>512, ...$), and used to create bounds for each $h^i$. We repeat this for all decision trees in the forest, and based on these local decision boundaries, we compute tight ranges for each hyperparameter. This procedure is illustrated in \autoref{fig:rf}.




\subsubsection{Model Validation}
To validate that a random forest model can sufficiently learn the regression problem formulated in section \ref{sec:model:random-forest}, we run \textit{k}-fold, \textit{k=10}, cross-validation with the $R$-squared correlation metric (a value between 0 and 1, where 0 represents no correlation and 1 represents perfect correlation) \cite{COLINCAMERON1997329} on a held-out dataset of each use-case's model configurations $\mathcal{H}$ and performance metrics $\mathcal{P}$. The average $R$-squared score across all folds for the image classification use-case was 0.89, and for the machine translation use-case, 0.91, showing that overall our model provides a reasonable fit for $\mathcal{H}\rightarrow\mathcal{P}$. Our full cross-validation results are available in \autoref{appendixE}\footnote{Please see \url{https://osf.io/vjq6m/?view\_only=e252c41da2d8446293b836adef40e443}}. 

\section{Case Studies}

We demonstrate two different use-cases of this tool, over a popular classification problem and a popular language generation problem for our user study in order to evaluate the efficacy of our features.\footnote{Training configurations are available in our supplemental materials.} The first use-case is image classification, which involves generating 244 models to classify images of handwritten digits found in the MNIST database \cite{mnist}, with the data evaluated by accuracy and training loss. 

The architecture of the neural networks used for this task is LeNet-5 \cite{lenet}. The hyperparameters adjusted in the study are as follows: \textit{beta 1} and \textit{beta 2} of the \textit{Adam} optimizer \cite{kingma2014method}, and \textit{learning rate}. To begin ascertaining the behavior of $\mathcal{H}\rightarrow\mathcal{P}$, we leveraged adaptive grid search.

The second use-case is machine translation -- the task of translating strings from one language to another, typically evaluated with the BLEU \cite{BLEU} metric. 
The machine translation use-case involves generating 205 models with the transformer \cite{transformer} architecture. These models were trained using the JoeyNMT framework \cite{kreutzer-etal-2019-joey} on a subset of en-de examples from WMT20 Chat Translation task\footnote{\url{http://www.statmt.org/wmt20/chat-task.html}}.



\section{Evaluation}
\label{sec:evaluation}

To evaluate our tool we conducted an IRB-approved user study initialized with the two use-cases mentioned above.
We used A/B testing to validate the predictive features of our dashboard based on a comprehension question posed to the participants. 

\subsection{Methodology}

We conducted this study entirely online with visual conferencing software. 
Participants who have experience using deep learning models were selected based on a pre-interview questionnaire filled out by interested parties. 
All interviews lasted between 45-75 minutes, and each interview was divided into three phases:
\begin{enumerate}
    \item A tutorial, in which we walked through all the primary features of the system.
    \item A comprehension check\footnote{Please see \url{https://osf.io/vjq6m/?view\_only=e252c41da2d8446293b836adef40e443} for a list of questions asked to interviewees} for a dashboard \textbf{without} highlighting suggestive features or importance scores described in \autoref{sec:predictive} (hereby known as dashboard A).
    \item A comprehension check for a dashboard \textbf{with} highlighting suggestive features or importance scores (hereby known as dashboard B). 
\end{enumerate}

After each comprehension check (e.g., asking which hyperparameter appears to have the most impact on a certain test metric, given the visualization data), participants completed a post-exploration questionnaire, in which they answered questions regarding their opinions on the validity and confidence they have in the predictive features (i.e., hyperparameter importance and predicted hyperparameter bounds). 
To diminish the effect of inherent differences in presenting inference features as discussed in \autoref{sec:predictive} on different datasets, the dashboard condition and use-cases were counterbalanced across all participants.

\subsection{Results}

We describe the strengths and limitations of our tool after evaluating it with users in our study. We discuss the efficacy of our tool, and frame it in the context of the use-cases users were given. 

An overall assessment of our tool is that it increases user ability to find optimal hyperparameter configurations with reasonably high confidence using the predictive features themselves. Based on feedback, our system is effective in guiding users towards narrowing the search space for future models; their reported confidence\footnote{An average score of 7.75 out of 10 in quantifying user confidence in the predictive features} in their decision-making process seems to validate the bounds suggested by the model. 

For the image classification models (presented as dashboard A), P1 and P2, who did not have access to our inferential features, were able to determine the most important hyperparameters and their values in the given space. However, based on the post-interview questionnaire response, P1's and P2's confidence in their choice was lower than that of P3 and P4, who had access to the inferential features when presented dashboard A. 

For the machine translation models (dashboard B), P3 and P4, who did not have access to the inferential features, put in significant manual effort to determine the most important hyperparameters and their values. P3 cites that "it is difficult to come to a conclusion [w.r.t. assessing hyperparameter impact] as the spread is so uniform." Meanwhile, P1 and P2, who had access to the inferential features when presented dashboard B, were able to cautiously confirm that the hyperparameter ranges highlighted by the tool would result in more optimal performance. This demonstrates that even with more complex and varied datasets, participants were still optimistic\footnote{An average score of 8.34 out of 10 in quantifying the efficacy of the tool in hyperparameter search} about the effectiveness of the implemented inferential features. 

With regard to improvements in the predictive components of the dashboard, participants wanted some form of validation of the predictive model. This was explicitly requested by P4, and similar sentiments were shared by P1 and P3. To address this, we updated the user interface to expose the $R$-squared score and accuracy of the random forest used to predict model performance, which provides a measurable metric for users to take into consideration when evaluating the automated features. 

\section{Conclusion and Future Work}

In this paper, we present a system for the recommendation of optimal hyperparameter values, embedded within a streamlined visual analytics tool. This system: (a) recommends optimal ranges for hyperparameters, and (b) ranks hyperparameters by their estimated impact on given performance metrics.

In a preliminary study with four users, we find that users find our automated guidance features intuitive and preferable to existing hyperparameter tuning techniques. Our participants indicate that narrowing down the hyperparameter search space, and leaving room for human choices, is able to optimize future iterations of model tuning.

In the future, we plan to have a more thorough study of the predictive methods, including more varied domains and model configurations (\textit{e.g.} how few models can the user provide before the suggestions become unhelpful?), for increasing confidence in the generalizability of our techniques. Furthermore, we plan to extend our interface to provide predictions for completely unseen hyperparameter configurations, although it would require more thought into the predictive model. Nevertheless, our system provides a launchpad for hyperparameter optimization using visual analytics.

\bibliographystyle{eg-alpha-doi}
\bibliography{egbibsample}

\onecolumn

\newpage
\appendix
\section{Pre-interview Survey}

\begin{table}[!htb]
    \begin{minipage}{0.5\linewidth}
        \centering
        \begin{tabular}{c c} 
            \hline
            How long have you \\worked with deep learning tools  \\ [0.5ex] 
            \hline\hline
            <1 year &  0  \\ 
            \hline
            1-2 years   & 0  \\
            \hline
            3-5 years   & 3  \\
            \hline
            6-7 years    & 1  \\ [1ex] 
            \hline
        \end{tabular}
    \end{minipage}
    \begin{minipage}{0.5\linewidth}
        \centering
        \begin{tabular}{c c} 
            \hline
            How would you rate your \\ area of expertise in deep learning?  \\ [0.5ex] 
            \hline\hline
            Novice &  0 \\
            \hline
            Intermediate & 1      \\
            \hline
            Advanced   & 2   \\
            \hline
            Expert    & 1    \\ [1ex] 
            \hline
        \end{tabular}
    \end{minipage}
\end{table}

\begin{table}[!htb]
    \begin{minipage}{0.5\linewidth}
        \centering
        \begin{tabular}{c c} 
            \hline
             How would you rate \\your expertise in finding the\\ best hyperparameters for a deep \\learning task in your \\area of expertise? \\ [0.5ex] 
            \hline\hline
            Novice &  0 \\
            \hline
            Intermediate & 2      \\
            \hline
            Advanced   & 2   \\
            \hline
            Expert    & 0    \\ [1ex] 
            \hline
        \end{tabular}
    \end{minipage}
    \begin{minipage}{0.5\linewidth}
        \centering
        \begin{tabular}{c c} 
            \hline
            How many projects have you\\ worked on that required fine-tuning \\hyperparameters?  \\ [0.5ex] 
            \hline\hline
            0 &  0 \\
            \hline
            1 - 2 & 1      \\
            \hline
            3 - 5   & 1   \\
            \hline
            6 - 10 &  1 \\
            \hline
            11 - 20 & 1      \\
            \hline
            21+  &  0   \\ [1ex] 
            \hline
        \end{tabular}
    \end{minipage}
\end{table}

\section{Interview Questions}
\begin{itemize}
    \item Could you tell me what the hyperparameters are in the dashboard?
    \item Could you tell me what the result statistics are in the dashboard?
    \item What set of hyperparameter values do you think the optimal model should be constructed of?
    \item What are the hyperparameter settings for the top 3 best performing models (“best performing” up to your interpretation)?
    \item Which hyperparameter do you think has the most impact on the test metrics? 
    \item Which hyperparameter(s) seem the most impactful to model performance and what common values do they take?
    \item In this set of experiments, do you see that hyperparameters are invariant to each other (Do they operate independently of each other, or are they very intertwined)?
    \item What hyperparameter configuration would you try next?
\end{itemize}

\section{Supplementary Visuals}
\vskip5mm
\begin{figure}[hbt!]
\centering
\includegraphics[width=.36\textwidth]{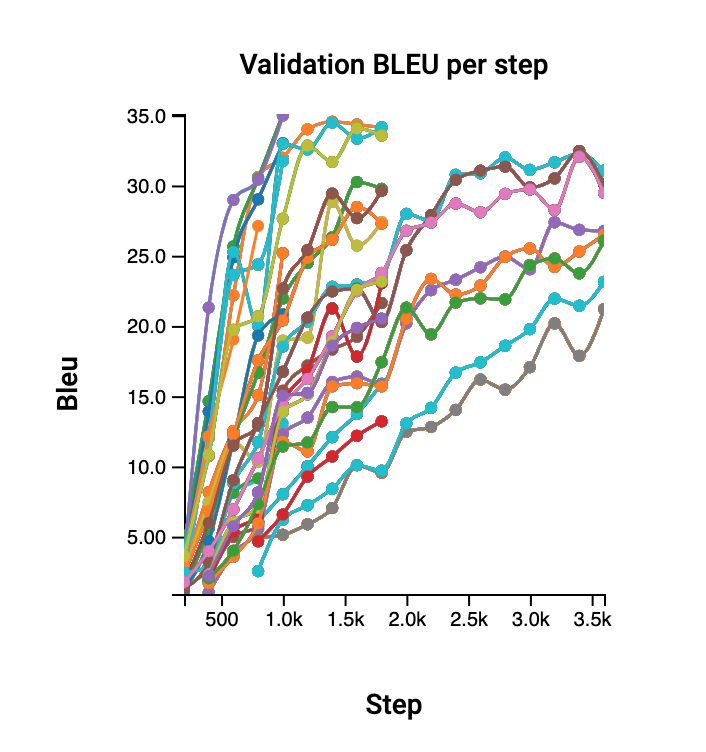}\hfill
\includegraphics[width=.416\textwidth]{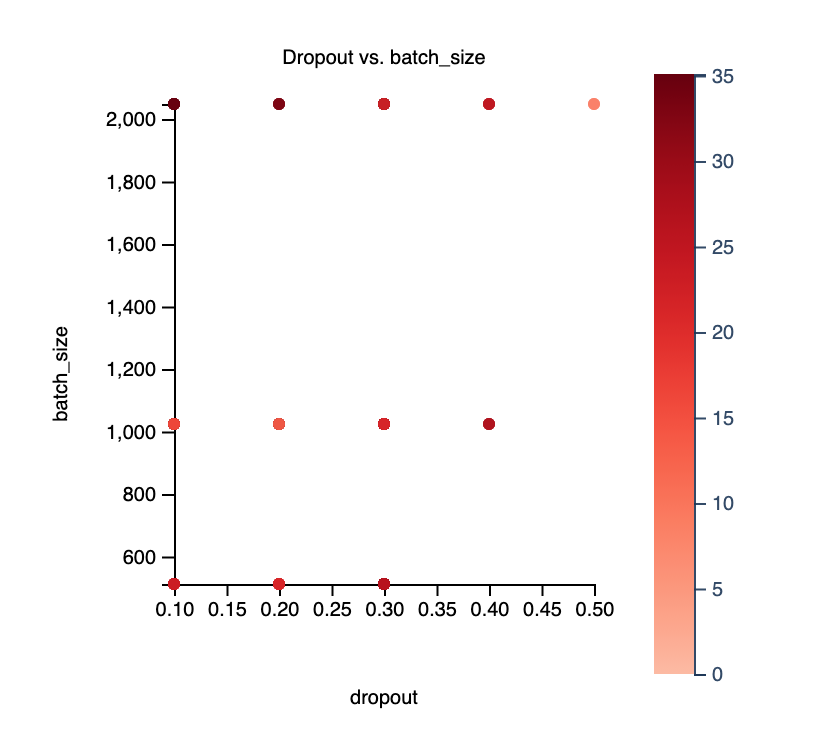}\hfill
\caption{Supplementary Visual Plots}
\label{fig:supplementary}
\end{figure}

\newpage
\section{Hyperparameter Search}
To initialize dashboards for study, we generated models with two use cases, counterbalancing features and dashboards in the study. This section in the appendix describes how we created configurations to experiment with.
\subsection{Image Classification}
\begin{table}[!htb]
    \begin{minipage}{\linewidth}
        \centering
        \begin{tabular}{ |p{2cm}||p{1cm}|p{1cm}|p{1.1cm}|  }
        \hline
        \multicolumn{4}{|c|}{Range for Image Classification Case} \\
        \hline
        Hyperparameter & Min & Max & Intervals\\
        \hline
        Beta 1 & 0.5 & 0.9 & 2\\        
        Beta 2 & 0.9 & 0.999 & 2\\
        Learning Rate & 1e-6 & 1 & 3\\
        \hline
        \end{tabular}
        \vskip0.5cm
        \caption{Preset range of hyperparameters. The scope of each hyperparameter narrows each time a grid search over the previous range is completed. The new grid centers on the coordinate from the previous grid yielding the best result, optimizing sequentially.}
        \label{hyp_range}
    \end{minipage}
\end{table}

For the image classification task, we use adaptive grid search, setting the initial range of the hyperparameters to be as shown in Table \ref{hyp_range}, and sequentially optimizing as the model trains. A list of values in a hyperparameter dimension bounded by \textit{Min} and \textit{Max} is displayed. \textit{Intervals} evenly divides the bounds specified, such that each number in the resulting list is an increment of (\textit{Max} - \textit{Min}) / (\textit{Intervals} - 1). A total of 244 models were generated.

\subsection{Machine Translation}

For the machine translation task, we use naive grid search, setting the parameter lists ahead of time as shown in Table \ref{tab:wmt}. Several models had to be early stopped due to incompatibility of certain hyperparameter values or resource constraints. A total of 205 models were generated.

\begin{table}[!htb]
    \begin{minipage}{\linewidth}
        \centering
        \begin{tabular}{ |p{2.85cm}||p{3.5cm}|  }
        \hline
        \multicolumn{2}{|c|}{Range for Machine Translation Case} \\
        \hline
        Hyperparameter & Min\\
        \hline
        \# Encoder Heads & 8, 16 \\
        \# Decoder Heads & 8, 16 \\
        Dropout & 0.1, 0.2, 0.3, 0.4, 0.5 \\
        Encoder hidden size & 128, 256, 512, 1024, 2048 \\
        Decoder hidden size & 128, 256, 512, 1024, 2048 \\
        Batch size & 512, 1024, 2048 \\
        \hline
        \end{tabular}
        \vskip0.5cm
        \caption{Hyperparameter lists for Machine Translation case, constructed for naive grid search}
        \label{tab:wmt}
    \end{minipage}
\end{table}

\newpage
\section{Cross-Validation Results}\label{appendixE}

Through \textit{k}-fold cross-validation\footnote{We use $R^2$ scoring as the scoring criterion because the performance metrics in these use-cases span continuous ranges.} on a held-out dataset of the provided model configurations, we find that our random forest model is able to accurately predict effective hyperparameter values in both of our target use-cases\footnote{For the Image Classification case, model configurations were generated serially using an adaptive grid search algorithm}.

\begin{table}[!htb]
    \begin{minipage}{\linewidth}
        \centering
        \begin{tabular}{ |p{2cm}||p{3cm}|p{3cm}| }
        \hline
        \multicolumn{3}{|c|}{Range for Machine Translation Case} \\
        \hline
        Fold Number & Machine Translation (modeling predictability on BLEU) & Image Classification (modeling predictability on Accuracy) \\
        \hline
        1 & 1.00 & 1.00 \\
        2 & 0.68 & 0.99 \\
        3 & 0.94 & 0.9 \\
        4 & 0.98 & 0.91 \\
        5 & 0.95 & 0.96 \\
        6 & 0.86 & 0.86 \\
        7 & 0.87 & 0.99 \\
        8 & 0.88 & 0.7 \\
        9 & 1.00 & 0.91 \\
        10 & 0.83 & 0.72 \\
        Average & 0.91 & 0.89 \\
        \hline
        \end{tabular}
        \vskip0.5cm
        \caption{Cross-Validation Scores to model $\mathcal{H}\rightarrow\mathcal{P}$ (for each dashboard used in the study). A 0.4/0.6 train/test split was used for 10-fold cross-validation on held-out test sets. MT had 82 model configurations to train on and 123 held-out models. IC had 97 model configurations to train on and 147 held-out models.}
        \vskip2cm
        \label{cross_validation}
    \end{minipage}
\end{table}

\end{document}